\documentclass[
reprint,
superscriptaddress,
amsmath,
amssymb,
aps,
prb,
longbibliography,
noeprint
]{revtex4-2}

\usepackage{epsfig}
\usepackage{dcolumn}
\usepackage{bm}
\usepackage{color}
\usepackage{graphicx}
\usepackage{bbold}

\begin{document}

\title{Quantum phase transitions in quantum Hall and other topological systems: role of the Planckian time}

\author{Andrey Rogachev} 
\affiliation {Department of Physics and Astronomy, University of Utah, Salt Lake City 84093, USA}

\date{\today}

\begin{abstract}
Transformations between the plateau states of the quantum Hall effect (QHE) are an archetypical example of quantum phase transitions (QPTs) between phases with non-trivial topological order \cite{WeiTsui88,EngleTsui90,KochvKlitzing92,ShaharTsui97}.  These transitions appear to be well-described by the single-particle network theories \cite{Huckestein95,DresselhausGruzberg21}. The long-standing problem with this approach is that it does not account for Coulomb interactions. In this paper, we show that experimental data in the quantum critical regime for both integer and fractional QHEs can be quantitatively explained by the recently developed phenomenological model of QPTs in interacting systems \cite{Rogachev23micro}.  This model assumes that all effects of interactions are contained in the life-time of fluctuations as set by the Planckian time $\tau_P=\hbar/k_BT$.  The dephasing length is taken as the distance traveled by a non-interacting particle along the bulk edge state over this time. We show that the model also provides quantitative description of QPTs between the ground states of anomalous QHE and axion and Chern insulators \cite{Wu_NC20,Liu 20}. These analyzed systems are connected in that the QPTs occur via quantum percolation. Combining the presented results with the results of two companion papers \cite{RogachevDavenport23,Rogachev23micro}, we conclude that the Planckian time is the encompassing characteristic of QPTs in interacting systems, independent of dimensionality and microscopic physics. 
\end{abstract}

\maketitle
Quantum phase transitions between the plateaus of quantum Hall states have been intensively studied since the discovery of the integer quantum Hall effect (IQHE). This was one of the first systems where the finite size scaling analysis was used to detect QPTs and extract values of critical exponents \cite{SondhiShahar97}. 

In heterostructures used to study QHE, charged donors randomly located in the remote doping layer produce a smooth electrostatic potential in the plane of the 2d electron gas. The degenerate electrons fill this potential profile and produce a disordered network illustrated in Fig. 1. The beige color marks the “puddles” filled by electrons and the blue color the unoccupied areas corresponding to the “hills” of the electrostatic potential. The current carrying states propagate along the equipotential lines separating the puddles and hills. 

\begin{figure}[b]
\centering
 \includegraphics[width= 0.6\columnwidth]{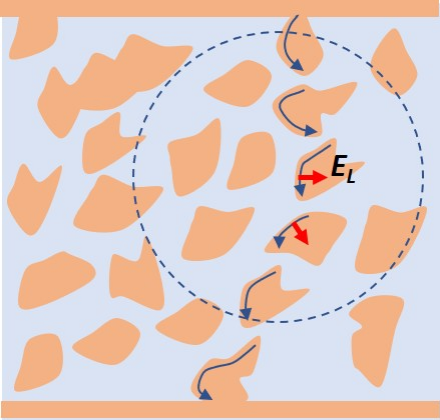}
 \caption{\textbf{Schematic picture of the quantum percolation in IQHE.} 
The solid arrowed line indicate edge modes on the border of electron puddles.  The red arrows indicate the electrical field $E_L$ and the dashed circle the area within the dephasing length $L_{\varphi}$.}
 \end{figure}

The QPT occurs when the system reaches the percolation threshold.  A recent series of exceptionally careful studies of IQHE in AlGaAs heterostructures  \cite{LiTsui05,LiTsui09,LiTsui10,LiTsui03} has established that the transition mechanism is determined by the size of a typical puddle, $L_p$, as compared to the dephasing length, defined as $L_\varphi \sim 1/T^{1/z}$, where $z$  is the dynamical critical exponent. In the so-called universal regime, when $L_p\ll L_\varphi$, electrons propagate along the border of the puddles with intermittent coherent tunneling from one puddle to another. This process of quantum percolation illustrated in Fig. 1 is the essence of the Chalker-Coddington network model \cite{ChalkerCoddington88} and its generalizations \cite{JainKivelsonTrivedi,LeeWangKivelson,Huckestein95,DresselhausGruzberg21}.  Numerical studies predict that in this regime, the correlation length exponent is in the range $\nu$=2.3-2.65 \cite{Huckestein95,DresselhausGruzberg21}; this value agrees with majority of experiments. The second non-universal regime corresponds to $L_p\sim L_\varphi$.  In this case, one expects the scaling behavior with a non-universal exponent $\nu$ \cite{LiTsui10,DodooAmoo14} or no scaling at all. 

\begin{figure*}[tbph]
\centering
 \includegraphics[width= 0.9\textwidth]{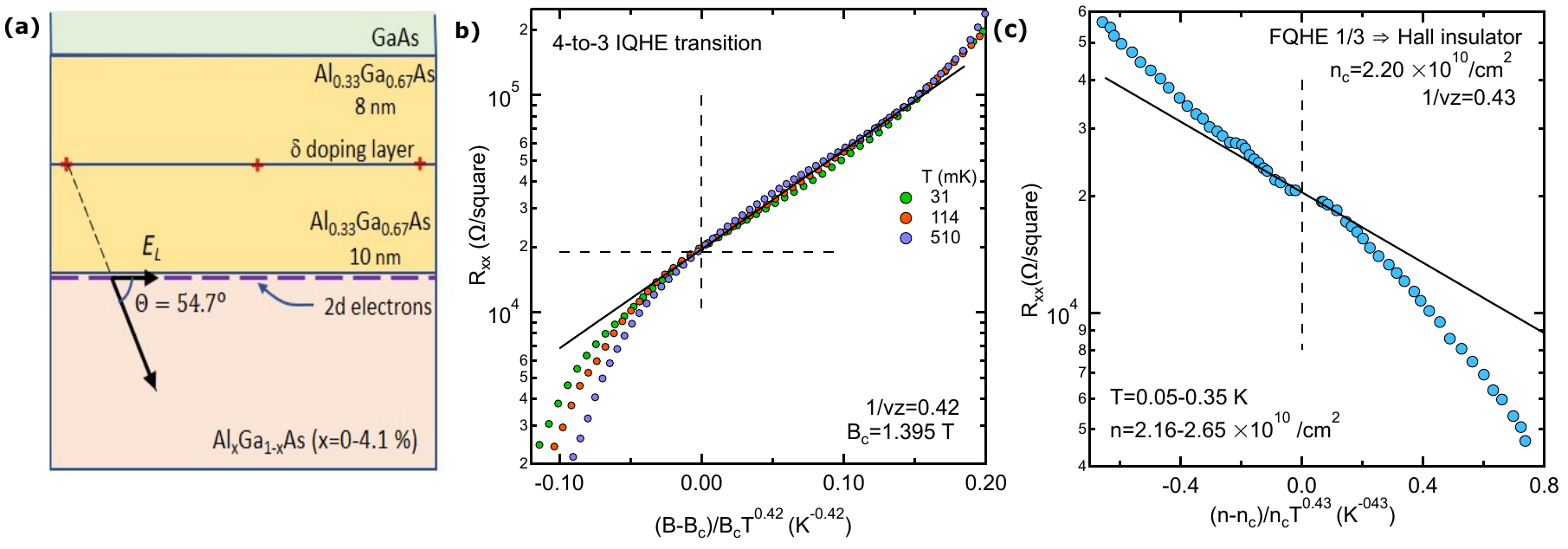}
 \caption{\textbf{Quantum phase transition in Quantum Hall systems}. (a-b) Integer QHE: (a) The sequence structure of the samples studied in \cite{LiTsui05,LiTsui09,LiTsui10,LiTsui03}. (b) The resistance $R_{xx}$ of the top-most Landau level versus the scaled magnetic field for the sample with $x=0.85\%$ (see text for details).(c) Fractional QHE \cite{WongTrivedi93}: Resistance $R_{xx}$ versus the scaled concentration of carriers.}
 \end{figure*} 

Despite their success, the network models have a major flaw: they do not account for Coulomb interactions. Significant efforts have been made over past thirty years to resolve this problem \cite{LeeWang96,HuckensteinBackhaus99,PruiskenBurmistrov07,OswaldRomer17,KumarRghu22}, but it is still fair to say, following \cite{GirvinYangBook}, that agreement between experiments and the network models, based on the single-particle physics and ignoring Coulomb interactions, is indeed “something of a mystery”. 

 In this paper, we propose the resolution to this problem (albeit only on a phenomenological level) by invoking the concept of Planckian time \cite{SachdevKeimer11,HartnollMackenzie22}. We analyze the QHE systems using the recently developed model of QPTs \cite{Rogachev23micro}, which proposes that conductivity varies across the transition according to the generic exponential relation 
\begin{equation}
\sigma(T)=\frac{e^2}{\hbar L_\varphi^{d-2}}g_c\exp{\left(\frac{y-y_c}{y_c}\left(\frac{L_\varphi(T)}{L_0}\right)^{1/\nu}\right).}
\end{equation}

Here, $y$ is the non-thermal parameter driving the transition and $y_s$ is its critical value. The model is applicable only in the quantum critical regime defined by condition that the dephasing length is smaller than the correlation length, $L_\varphi<\xi \sim \left|y-y_c\right|^{-\nu}$. The model has the key parameter, $L_0$, the microscopic seeding scale of the real space renormalization group. For many systems, it is fairly clear what $L_0$ is, so one can make assumptions about the dependence of  $L_{\varphi}(T)$ on system properties, extract experimental $L_0$, check if it matches the relevant scale, and thus gain some understanding of microscopic processes governing the transition. 

We found that in all interacting systems tested in \cite{Rogachev23micro,RogachevDavenport23}, agreement with the experiment occurs if the dephasing length is chosen as the distance travelled by a non-interacting carrier (or excitation) over the Planckian time, $\tau_P$=$\hbar/k_BT$.  Here, we show that this approach also works for topological systems where the QPT occurs via quantum percolation.  
\linebreak

\noindent\textbf{Quantum phase transitions: IQHE and FQHE.}

\textit{Integer QHE.} For our analysis of the integer QHE, we have chosen the experimental results presented in \cite{LiTsui05,LiTsui09,LiTsui10}. The sequence structure of the samples is described in \cite{LiTsui03} and depicted in Fig. 2a. The 2d electron gas resides at the interface between Al$_{0.33}$Ga$_{0.67}$As and Al$_x$Ga$_{1-x}$As on the side of Al$_x$Ga$_{1-x}$As. Aluminum content of this layer was varied in the range  $x$=0-4.1$\%$.  When an Al atom substitutes Ga atom in the lattice, it produces strong atomic-scale scattering potential. At high magnetic fields, the fractional QHE was observed though not studied in \cite{LiTsui05}.

We start with the analysis of the data for the 4-to-3 plateau-to-plateau transition in a sample with $x$=0.85$\%$. We traced the data for $R_{xx}\left(B,T\right)$ and $R_{xy}\left(B,T\right)$ variations shown in Fig.1c,d of \cite{LiTsui05}. We then used the procedure described in \cite{ShaharTsui97} to obtain the resistance $R_{xx}$ of the topmost Landau level.  Figure 2(b) shows that this resistance can be scaled using the values of $B_c$=1.395 T and $\nu z$$\approx$2.38 reported in \cite{LiTsui05}. Near the transition, the data can be approximated by the exponential dependence $y$=$a_1\exp \left(a_2x\right)$ with $a_1$=19500 $\Omega/\square$ and $a_2$=10.5 K$^{-0.42}$. 

 The value of the dynamical exponent for this sample is $z$$\approx$1. It was independently determined from the measurements of the samples with different sizes \cite{LiTsui09}. The well-accepted view in QHE literature is that the dynamical exponent is determined by some unspecified inelastic process.  However, there are also long-standing arguments questioning this interpretation.  First of all, the exponent $z$=1 disagrees with the non-interacting model, which predicts $z$=2 (page 380 in \cite{Huckestein95}). Moreover, frequency-dependent measurements on many systems show that the transition from the frequency-dominated regime to the temperature-dominated regime is determined by the condition $\hbar\omega\approx k_BT$ \cite{EngelTsui93,HohlsKuchar02}.  As pointed out early on (page 327 in \cite{SondhiShahar97}), this behavior indicates that dephasing comes not from the system-specific microscopic scattering but rather is a universal signature of the quantum critical regime of an interacting system. This is what we take as a starting assumption in the analysis of the data presented in Fig.2b. 

We set the time scale of the fluctuations to the Planckian time $\tau_P$=$\hbar/k_BT$.  Then, according to our general conjecture described in \cite{Rogachev23micro}, the dephasing length is given by the distance traveled by a non-interacting particle over $\tau_P$. From the semiclassical picture of quantum percolation, this length is determined by the drift of a particle along the equipotential lines interrupted by tunneling events between neighboring electrostatic contours as shown in Fig.1. Since these tunneling events are essentially instantaneous, the propagation is dominated by the drift and, to the first approximation, the dephasing length can be given as $L_\varphi\approx v_{dr}\hbar/k_BT$. Notice that this simple relation provides the needed dynamical exponent $z$=1. We estimate the drift velocity from the semiclassical relation $v_{dr}$=$\vec{E}\times\vec{B}/B^2$=$E_L/B$, where $E_L$  is the in-plane component of the typical electrical field at the edge of a puddle and $B\approx B_c$ in the quantum critical regime.  Then, for IQHE, the generic scaling Eq.1 takes the specific form 
\begin{equation}
R_{xx}=R_c \exp\left(\ \left(\frac{\hbar E_L}{k_BB_cL_0}\right)^{1/\nu}\frac{B-B_c}{B_c}\ \frac{1}{T^{1/\nu}}\right).
\end{equation}

The minimal seeding scale of the QPT in IQHE is expected to be set by the magnetic length $l_m$=$\left(\hbar/eB_c\right)^{1/2}$.  In the present case, it is $l_m\approx22$ nm. Using this value for $L_0$ and experimental values of the exponent $\nu=2.38$ and coefficient $a_2$, we estimate the value of the in-plane electrical field as $E_L\approx1.0\times{10}^6$ V/m.  

Now we need to get an independent estimate of $E_L$. Notice that this field is zero when the smooth disorder potential is either empty or completely filled by electrons; the percolation threshold is reached somewhere in between. To proceed further, we choose to ignore the electron screening (see discussion below) and take as an estimate of $E_L$ the maximum value of the in-plane field produced by a single donor with the charge +e located in the $\delta$-doping layer, 10 nm away from 2d electron gas. This field is shown in Fig 2(a) and Fig.1. We find from the geometry of the sample and Coulomb’s law that  $E_L\approx4.0\times{10}^5$ V/m. This value agrees with experimental $E_L$ in order of magnitude.  

Let us now check if our model explains other observations reported in \cite{LiTsui05,LiTsui09,LiTsui10}. We have traced data for $R_{xx}\left(B,T\right)$ and $R_{xy}\left(B,T\right)$ dependences for the 2-to-3 transition from Fig. 1(a,b) of \cite{LiTsui05}, and repeated the analysis. We have found that the data can be scaled using the critical field $B_c=1.90$ T and universal exponent $\nu z$=$2.38$.  The in-plane electrical field for this transition is $E_L\approx8.5 \times{10}^5$ V/m. This value confirms the assumptions of the model.

 At low temperatures, samples with low concentration of Al impurities reveal the power-law temperature dependence for the 3-to-4 transition, $\left(dR_{xy}/dB\right)|_{B_c}$=$a_0T^{-\kappa}$, with the universal value of $\kappa\approx0.42$. ( The exponents $\kappa$ and $p$ used in the quantum Hall literature relate to $z$ and $\nu$ as $\kappa$=$1/\nu z$ and $p$=$2z$.) 
We have extracted the value of the proportionality constant $a_0$ from Fig. 2 of Ref.\cite{LiTsui10} and found that it varies very little, within 10 $\%$, for samples with $x$=0, 0.21, and 0.85$\%$. This is what we expect from our model. It assumes that the coefficient $a_0$ depends only on two sample-dependent parameters, $B_c$ and $E_L$. The samples have similar carrier concentrations (see Table 1 in \cite{LiTsui05}) and, hence, similar values of $B_c$. The in-plane field $E_L$ is determined by the thickness of the spacer layer, which is the same for all samples.  Let us notice that the similar values of $a_0$ for three samples with different concentration of the short-range impurities suggest that while these impurities define the size of the puddles (as was concluded in \cite{LiTsui10}), they are of secondary importance for $L_\varphi$, which is determined mostly by the drift along the puddles’ perimeters. 

In Ref.\cite{LiTsui09}, the study of the heterostructure with $x$=0.85$\%$ was extended to the measurements of a series of samples with the varied size.  It was found that at low temperatures the  $\left(dR_{xy}/dB\right)|_{B_c}$=$a_0 T^{-0.42}$ variation crosses over to the temperature-independent behavior. The “knee” point between two regimes was defined as the saturation temperature, $T_s$, and it was found to scale with the width of the samples as  $W$=$b_sT_s^{-1}$.  From Fig.3 of \cite{LiTsui09} we found that $b_s$=$2.9 \times{10}^{-5}$ (m$\times$K). The authors of the work pointed out that the saturation occurs because the dephasing length becomes comparable with the smallest sample size (the width in the studied case).  The dephasing length in our model depends on temperature as $L_\varphi\approx v_{dr}\hbar/k_BT$=$b_\varphi T^{-1}$, with $b_\varphi\approx5.8\times{10}^{-6}$ (m$\times$K).  We interpret this length as a typical size of fluctuations, which appear randomly in all positions of the sample. At high temperatures, only fluctuations located on the edge of the sample will be affected by the edge presence. The crossover will be completed when the fluctuations originating on one edge of the sample reach the other. It is then reasonable to assume that the “knee” and $T_s$ correspond to the midpoint situation when $L_\varphi\approx W/2$ or smaller. With this assumption our model provides fairy good semi-quantitative description of the size effect.  

\textit{Fractional QHE (FQHE)}. Let us now consider the study of the fractional QHE reported in \cite{WongTrivedi93,JiangHannhs93}, which gives sufficient details of the sample structure to allow for a quantitative analysis of the QPT.  In this study, the 2d electron gas was separated by 80-nm-thick Al$_{0.3}$Ga$_{0.7}$As undoped spacer from Si-doped 30-nm-thick AlGaAs layer. From data presented in \cite{WongTrivedi93,JiangHannhs93},  we find that only a small portion of donors is ionized, hence we can assume that these ionized donors are located at a distance $d\approx80$ nm from 2d electron gas. The studied transition was between the $\nu$=1/3 Hall state and the Hall insulator state; it was driven by gate-induced change of the carrier concentration, $n$, at a fixed magnetic field $B$=2.91 T. We have traced the scaled data from Fig. 3(a) of \cite{WongTrivedi93} and plotted them in Fig 2(c) using $z\nu\approx2.3$, found in \cite{WongTrivedi93}. Near the critical concentration, the data can be approximated with the exponential fit $y$=$a_1\exp\left(-a_2x\right)$, with $a_1$=20500 $\Omega/\square$ and $a_2$=1.05$\pm0.15$ K$^{-0.43}$.  

As the starting point, we take the theoretical picture proposed in \cite{LeeWangKivelson} that, microscopically, the composite fermions propagate as undivided entities along the contours of electrostatic potential essentially in the same way as electrons and holes in the integer QHE, i.e. they repeat the processes shown in Fig.1. In addition, in agreement with our general conjecture \cite{Rogachev23micro}, we assume that in the quantum critical regime, the system is in an entangled, strongly-interacting state so that the composite fermions themselves are only elementary excitations living during the Planckian time $\tau_P$=$\hbar/k_BT$.  Since the drift velocity does not depend on the particle charge, the dephasing length for composite fermions is the same as the one for the integer QHE discussed above, $L_\varphi$=$E_L\hbar/k_BTB$, where $B$=2.91 T. The scaling equation is the same as Eq. 2, except that in the driving term, $B$ is replaced by $n$.

Similar to IQHE, we take $L_0$ equal to the magnetic length  $l_m$=$\left(\hbar/eB\right)^{0.5}\approx15$ nm. Then, using the experimental value of $a_2$, we find the in-plane experimental electrical field to be $E_L$=$6000\pm2000$ V/m. Also similar to IQHE, we get an independent estimate of $E_L$ as the maximum in-plane field produced by an ionized donor located at the distance $d$=80 nm away from 2d gas. We find $E_L\approx6500$ V/m, which matches closely our experimental value.  

The estimate of the typical in-plane electrical field is the obvious questionable step in our analysis. In fact, we just tried the simplest possible way of getting $E_L$ and it happened to work well. In reality, the effect of Coulomb interactions on IQHE presents a long-standing problem (see \cite{WernerOswald20} and references herein).  The latest numerical investigation of IQHE with the self-consistent Hartree-Fock approximation suggests that even the half-filled Landau level provides only poor screening when compared to the picture drawn from Thomas-Fermi-like approximations \cite{OswaldRomer17,RomerOswald}.  This work also concludes that “the IQH regime is dominated by many-particle physics that seems to act towards re-establishing the behavior expected for non-interacting electrons – as often assumed in early percolation-type models of IQHE effect.”  This statement seems to support our finding, but let us recall that our model does not neglect Coulomb interactions, rather it states that they are contained in the Planckian time.  From the success of the model, the following questions arise: how does the model relate to the microscopic picture found in \cite{WernerOswald20,OswaldRomer17,RomerOswald} and how do the Coulomb interactions affect the quantum fluctuations? If, as the theory of QPTs \cite{SondhiShahar97,SachdevKeimer11} suggests, the Planckian time is the shortest possible relaxation time, it is logical to expect that what carriers "see" in the quantum critical regime is just the static unscreened  Coulomb potential.   

\begin{figure}[b]
\centering
 \includegraphics[width= 0.8\columnwidth]{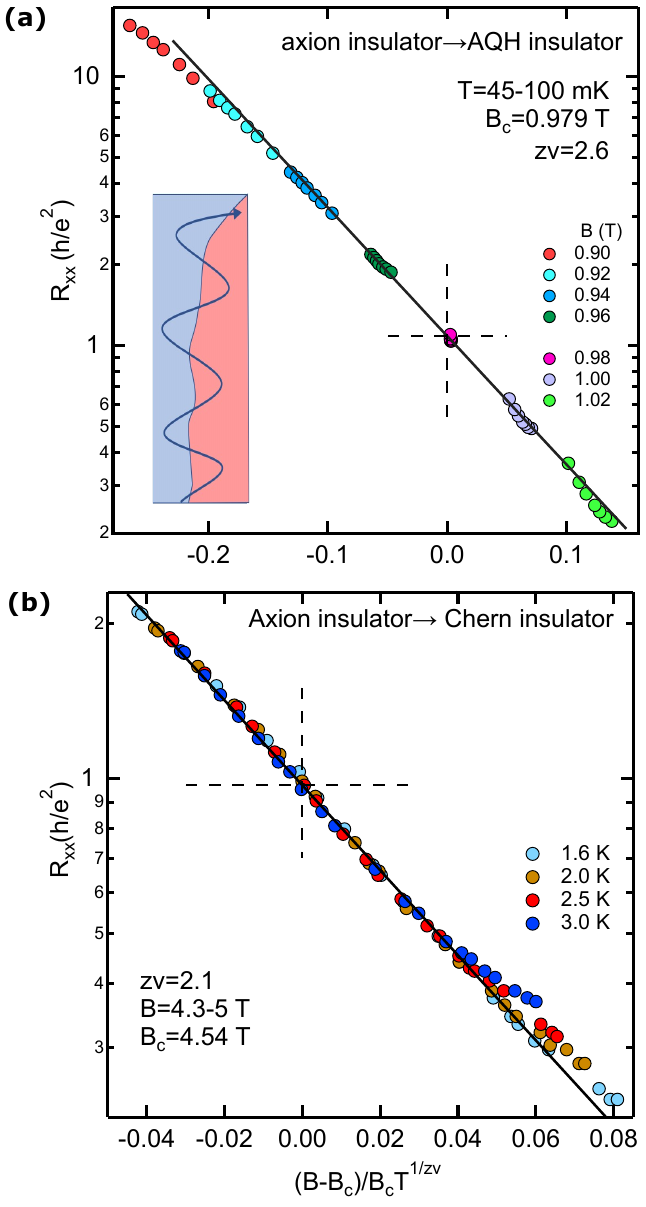}
 \caption{\textbf{Quantum phase transition in topological insulators.} 
(a) The resistance versus the scaled magnetic field for the thin film (Bi,Sb)$_2$Te$_3$ studied in \cite{Wu_NC20}. The inset shows the proposed edge mode propagating along the border of two magnetic domains with opposite magnetization. (b) The resistance versus the scaled magnetic field for thin film MnBi$_2$Te$_4$ studied in \cite{Liu 20}.}
\end{figure}

The model requires further confirmation from experiments. Let us mention few obstacles that complicate analysis of the existing data.  If filtering of cryostat measuring lines is not done properly, erroneous effects can be observed \cite{TamirShahar15}. Extraction of the dynamical exponent $z$ from non-linear conductance does not look reliable since it can be easily mixed up with heating effects \cite{GodlmanReview,KimRogachev16,RogachevSacepe20}. Scaling of the data can be accidental \cite{RogachevSacepe20}. The samples also need to be better characterized. It was mentioned in \cite{ShaharPHD95} that the early studies of QPTs in IQHE were performed on low mobility samples that “failed for various reasons in a way that resulted in a much poorer quality than expected”. If such failure comes from appearance of charged defects or migrated donors in the spacer layer, the estimate of $E_L$ obviously needs to be changed. 
\linebreak

\noindent\textbf{QPT between the states of topological insulators}

Let us now discuss two examples of contemporary topological systems undergoing QPTs. The first example is the QPT from the anomalous-quantum-Hall state to the axion insulator state \cite{Wu_NC20}. The heterostructure used in this experiment is based on a thin film of (Bi,Sb)$_2$Te$_3$ sandwiched between Cr- and V-doped  magnetic layers of the same material. Transition is driven by perpendicular magnetic field, which flips magnetic domains in two magnetic layers. We have traced the raw data from Fig. 2(b) of this work. Then we have scaled them using the values $B_c$ and $z\nu$=2.6 determined in \cite{Wu_NC20}; the scaled data are presented in Fig 3(a). The second example is the magnetic-field driven QPT between the axion insulator and the Chern insulator in a thin film of the antiferromagnetic topological insulator MnBi$_2$Te$_4$ \cite{Liu 20}. The scaling plot was generated using the data shown in Fig. 4(d) of this work and is presented in Fig. 3(b). The value of $zv$=2.1 is the same as in the original work.

As Fig. 3 attests, the scaled data for topological insulators follow the exponential dependence across the critical point of the QPT; this behavior suggests that they can be explained using the generic Eq. 1. For reference, the coefficients in the fitting equation $y$=$a_1\exp\left(-a_2x\right)$ are:  $a_1$=1.1  and $a_2$=10.5 K$^{-1/2.6}$ for panel (a), and $a_1$=0.97 and $a_2$=19 K$^{-1/2.1}$ for panel (b). 

In Refs. \cite{Wu_NC20,Liu 20} and in theoretical proposals \cite{WangZhang14,LiuZhang16}, the discussion of the quantum critical regime in topological insulators proceeds essentially by mimicking the Chalker-Coddinton network model. Indeed, the picture of the sub-micrometer magnetic domains visualized using a scanning nanoSQUID probe in Cr-doped (Bi,Sb)$_2$Te$_3$ \cite{LachmanZheldov16} is reminiscent of the array of filled and empty zones in the random potential of the network models.

 For both systems shown in Fig.3, the extracted values of the critical exponents suggest that the quantum percolation transition takes place and that the dynamical exponent  $z\approx1$. This value of $z$, and similarity with IQHE suggest that the dephasing length is set by the edge states propagating into the bulk of the systems along the walls of the magnetic domain structure over the Planckian time.

As a trial, we propose that electrons follow the semiclassical meandering trajectory shown in the inset to Fig.3(a). Then, the dephasing length is given as  $L_\varphi$=$v_{dr}\hbar/k_BT$ and the drift velocity $v_{dr}$ relates to the Dirac velocity of the TI Hamiltonian $v_D$ as $v_{dr}$=$a_3v_D$, where $a_3$ is a constant somewhat smaller than one. With these assumptions, the scaling equation becomes 
\begin{equation}
R_{xx}=R_c \exp\left(-\left(\frac{\hbar a_3 v_D}{k_BL_0}\right)^{1/\nu}\frac{B-B_c}{B_c} \frac{1}{T^{1/\nu}}\right).
\end{equation}
Using the experimental values of the coefficient $a_2$ and $v_D\approx8\times{10}^5$ m/s \cite{ZollnerFabian21} we find that the minimal seeding scale for axion-to-AQH transition in (Bi,Sb)$_2$Te$_3$ (Fig 3a)  is $L_0$=$a_3\times7.5$ nm. As expected, this length is much smaller than the submicron size of the domains visualized in \cite{LachmanZheldov16}. 

The value of $L_0$ corresponds to the minimum length scale on which the topological phase can exist. We have found that it is fairly well matched by the confinement length, $L_c$, defined by the condition that the energy level spacing in 2d area confined by $L_c$ is equal to the gap $E_b$ between the bulk energy bands. With $E_b$=0.28 eV \cite{LinderSudbo09}, we estimate $L_c\approx2\sqrt2\hbar v_D/E_b\approx3$ nm.   
 As a complimentary view of this length, let us notice that quantum fluctuations set by the uncertainty relation in the region confined by $L_c$ reach the bulk gap in the system. 

Using the same interacting model, we have analyzed axion-to-Chern transition in MnBi$_2$Te$_4$ (Fig. 3b).  With values of the Dirac velocity $v_D\approx4\times{10}^5$ m/s \cite{Trang_MnBiTe21} and the bulk gap $E_b$=0.16 eV \cite{Li_MnBiTe19}, we have found that similarly to the previous system, the minimal seeding scale of the QPT, $L_0$=$a_3\times6.3$ nm,  is close to the length scale set the level spacing, $L_c\approx4.5$ nm.

In summary, we have found that the phenomenological model incorporating the Planckian time provides a quantitative description of QPTs between various topological states of the 2d electron gas. All considered transitions belong to the universality class of quantum percolation. 

In two companion papers \cite{Rogachev23micro,RogachevDavenport23}, we have found that the model also works for QPTs in many other interacting systems. While the dephasing length is specific for each system, the Planckian time has emerged as the truly universal characteristic of the quantum critical regime. 

The important conclusion of our work is that the scaled data in the quantum critical regime carry information about microscopic processes and scales leading to QPTs. It is, therefore, highly desirable to apply modern numerical methods of the type used for IQHE in \cite{WernerOswald20,OswaldRomer17,RomerOswald} to analyze experimentally studied samples. The conclusion is, in fact, very general and certainly relevant to many other complex systems undergoing QPTs.

\bigskip
\noindent \textbf{ Acknowledgements}
 The research was supported by National Science Foundation under awards DMR1904221 and DMR2133014. The author thanks E. Mishchenko for valuable comments.

\end{document}